\documentclass[12pt,a4paper]{article}
\usepackage{epsfig}
\def\pslash{\rlap{\hspace{0.02cm}/}{p}}
\def\eslash{\rlap{\hspace{0.02cm}/}{e}}

\begin{document}
\title{Associated production of neutral toppion with
  a pair of heavy quarks in $\gamma\gamma$ collisions }

\author{Xuelei Wang$^{(a,b)}$, Bingzhong Li$^{(b)}$,Yueling Yang $^{(b)}$   \\
 {\small a: CCAST (World Laboratory) P.O. BOX 8730. B.J. 100080 P.R. China}\\
 {\small b: College of Physics and Information Engineering,}\\
\small{Henan Normal University, Xinxiang  453002. P.R.China}
\thanks{This work is supported by the National Natural Science
Foundation of China, the Excellent Youth Foundation of Henan
Scientific Committee, the Henan Innovation Project for University
Prominent Research Talents.}
\thanks{E-mail:wangxuelei@263.net}}
\maketitle
\begin{abstract}
\hspace{5mm}We have studied a neutral toppion production process
$\gamma\gamma\rightarrow f\overline{f}\Pi_{t}^{0}(f=t,b)$ in the
topcolor-assisted technicolor(TC2) model. We find that the cross
section of $\gamma\gamma\rightarrow t\overline{t}\Pi_{t}^{0}$ is
much larger than that of $\gamma\gamma\rightarrow
b\overline{b}\Pi_{t}^{0}$. On the other hand, the cross section
can be obviously enhanced with the increasing of c.m.energy. With
$\sqrt{s}=1600$ GeV, the cross section of $t\bar{t}\Pi_t^0$
production can reach the level of a few fb. The results show that
$\gamma\gamma\rightarrow t\bar{t}\Pi^0_t \rightarrow
t\bar{t}(t\bar{c})$ is the most ideal channel to detect neutral
toppion due to the clean SM background. With such sufficient
signals and clean background, neutral toppion could be detected at
TESLA with high c.m.energy.
\end {abstract}
\vspace{1.0cm} \noindent
 {\bf PACS number(s)}: 12.60Nz, 14.80.Mz,.15.LK, 14.65.Ha
\newpage
\noindent{\bf I. Introduction}

 At present the success of electroweak standard model  is
  well known and doubtness.
However, the mechanism of the electroweak symmetry breaking (EWSB)
is not yet quite understood. The Higgs particle that is assumed to
trigger the EWSB in the EWSM has not been found. In addition,there
are prominent problems of triviality and unnaturalness
\cite{y1,y2} in the Higgs section in the EWSB. Generally, it is
said that the present theory of EWSB is only valid up to a certain
energy scale $\Lambda$, and new physics beyond the EWSB will
become dominant above $\Lambda$. Possible new physics are
supersymmetry (SUSY) and dynamical EWSB mechanism concerning new
strong interactions, etc.

The advantage of dynamical EWSB is that it discards the elementary
scale field in the theory, so, it can completely avoid the
problems of triviality and unnaturalness. The simplest model
realizing this idea is the initial technicolor (TC) model proposed
independently by  Weinberg and Susskind \cite{y2,y3}. However,
such a simple model predicts a too large oblique correction
parameter S and has been refused by LEP data \cite{y4,y5}. In
order to overcome the shortcomings of the simplest model and
explain the large mass difference between the top quark and the
bottom quark, an important model called topcolor-assisted
technicolor (TC2) model is proposed by Hill \cite{y6} which is of
great value to be studied in the future high energy experiment.

In TC2 theory, the EWSB is driven mainly by technicolor
interactions. The extended technicolor gives contributions to all
ordinary quark and lepton masses including a very small part of
the top quark mass: $m_{t}^{'}=\varepsilon
m_{t}(0.03\leq\varepsilon\leq0.1)$ \cite{y6}. The topcolor
interaction also makes small contributions to the EWSB and gives
rise to the main sector of the top mass $(1-\varepsilon)m_{t}$.
One of the most general predictions of TC2 model is the existence
of three physical partical Pseudo-Goldstone Boson called toppion:
$\Pi_{t}^{\pm},\Pi_{t}^{0}$, which mass is in the range of
hundreds of GeV. The toppions can be regarded as the typical
feature of TC2 model. Thus, studying the possible signatures of
toppions and toppion contributions to some processes at the high
energy colliders is a good method to test TC2 model.

 The neutral toppion can be probed directly in
the decay processes $\pi_{t}^{0}\rightarrow \gamma\gamma, gg ,
\gamma z$ through an internal top quark loop, and the decay
processes $\pi_{t}^{0}\rightarrow \overline{t}c,b\overline{b},
t\overline{t}$ (if this is kinetically allowed). These possible
decay modes have been calculated in detail \cite{y7} and it shows
that the dominant decay mode is $\pi_{t}^{0}\rightarrow
\overline{t}c$ if $\pi_{t}^{0}\rightarrow \overline{t}t$ is not
allowed. The alternative way to probe toppions is to study some
toppion production processes. The future linear colliders(LC) will
provide an almost unique place to explore the toppion due to its
clean environment and high luminosity. Recently, we have studied
the neutral toppion production processes
 in high energy $e^{+}e^{-}$ and $e\gamma$ collision\cite{wang}, the studies provide
the feasible ways to detect toppion events and test TC2 model.

In this paper, we shall study an associated production of neutral toppion with
  a pair of quarks in $\gamma\gamma$ collisions. i.e.,
$\gamma\gamma\rightarrow f\overline{f}\Pi_{t}^{0}(f=t,b)$. The
advantage of photon colliders for some process has been
extensively explained in the literature \cite{y9,y10}. Some
similar processes have been thoroughly studied in the standard
model(SM)($\gamma\gamma\rightarrow t\overline{t}H$)\cite{tth-sm}
and in the minimal supersymmetric extension of the SM(MSSM)
($\gamma\gamma\longrightarrow t\overline{t}\phi$
$(\phi=h^{0},H^{0},A^{0})$)\cite{tth-mssm}. In the SM, the results
show that the cross section of $\gamma\gamma\rightarrow
t\overline{t}H$ is at the level of a few fb\cite{tth-sm}. In MSSM,
it is shown that when tan$\beta$ is not too large the associated
$h^{0}$  production is dominant, with the cross sction of 1.0 fb
or higher for phenomenologically favored values of the parameters.
The studies in Ref.\cite{he-h-j} have shown that $\gamma\gamma$
colision have the significient advantage to probe charged Higgs
boson and a polarized $\gamma\gamma$ collider can determine the
chirality of the Yukawa couplings of fermions with charged Higgs
boson via single charged Higgs boson production, and thus
discriminate models of new physics.

 This paper is organized as
follows. In set.II, we will present the calculations of the
production cross section of the process
$\gamma\gamma\longrightarrow f\overline{f}\pi_{t}^{0}$ $(f=t,b)$.
The results and conclusion will be shown in set.III.

\noindent{\bf II The calculation of the production cross section}

As it is known, the couplings of toppions to the three family
  fermions are non-universal and the toppions have large Yukawa couplings
  to the third generation. The coupling of $\Pi_{t}^{0}$
   to a pair of top or bottom quarks is proportion to the mass of quark
   and the explicit form can be written as\cite{hehongjian, y6}
\begin{equation}
-\frac{tan\beta}{v_{w}}[(1-\varepsilon)m_t\overline{t} r_{5}t
+m_b^*\overline{b}r_{5}b]
\end{equation}
Where tan$\beta=\sqrt{(\frac{v_{w}}{v_{t}})^{2}-1}$,$v_{w}=246$
 GeV is the EWSB scale and $v_{t}\simeq 60-100$ GeV  \cite{hehongjian} is
the toppion decay constant. $m_b^*\approx 6.6k$ is an instanton
induced b-quark mass, and $k\approx$ 1 to 0.1. With above
couplings, $\Pi^0_t$ can be produced associated with a pair of top
or bottom quarks in $\gamma\gamma$ collisions. The Feynman
diagrams for the process $\gamma(p_{1})\gamma(p_{2})\rightarrow
f(p_{3})\overline{f}(p_{4})\Pi_{t}^{0}(p_{5})$ $(f=t,b)$ are shown
in Fig.1 in which the cross diagrams with the interchange of the
two incoming photons are not shown.
\begin{figure}[h]
\begin{center}
\begin{picture}(150,100)(0,0)
\put(-100,-200){\epsfxsize125 mm \epsfbox{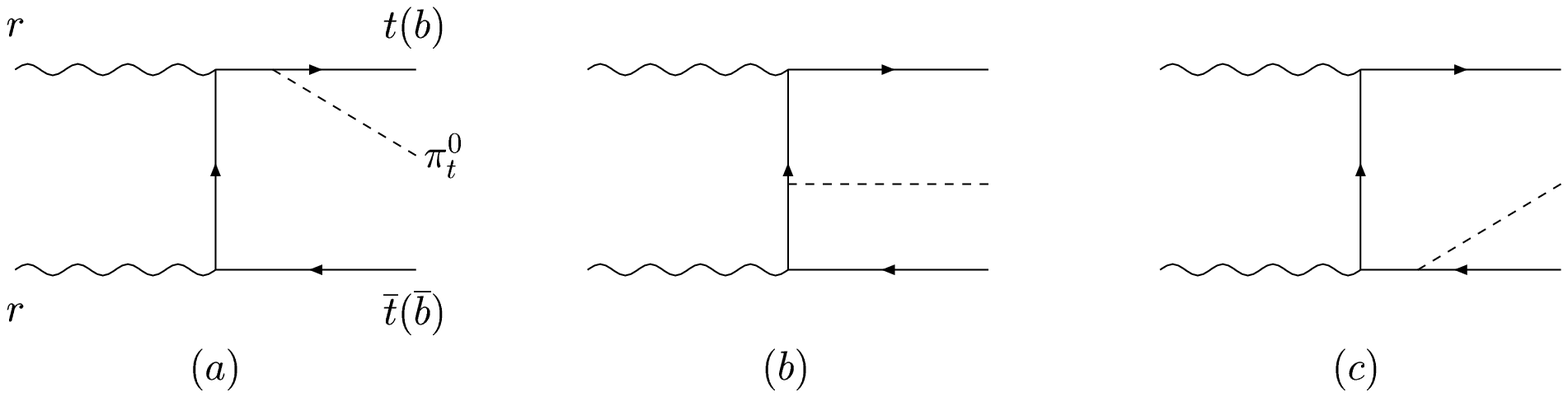}}
 \put(-60,-10){Fig.1 Feynman diagrams for
 $t\overline{t}(b\overline{b})\Pi^{0}_{t}$ production in $\gamma\gamma$-collisions}
\end{picture}
\end{center}
\end{figure}

The amplitudes for the process  are given by
\begin{eqnarray}
M^{(a)}_f&=&C_f\cdot G(p_{3}+p_{5},m_{f})G(p_{2}-p_{4},m_{f})\\
\nonumber& &
\overline{u}_{f}(p_{3})(\pslash_{3}+\pslash_{5}-m_{f})\eslash(p_{1})
(\pslash_{2}-\pslash_{4}-m_{f})\eslash(p_{2})\gamma_{5}v_{f}(p_{4})\\
\nonumber
 M^{(b)}_f&=&C_f\cdot G(p_{3}-p_{1},m_{f})G(p_{2}-p_{4},m_{f})\\
 \nonumber& &
\overline{u}_{f}(p_{3})\eslash(p_{1})(\pslash_{3}-\pslash_{1}+m_{f})
(\pslash_{2}-\pslash_{4}-m_{f})\eslash(p_{2})\gamma_{5}v_{f}(p_{4})\\
\nonumber
 M^{(c)}_f&=&-C_f\cdot G(p_{3}-p_{1},m_{f})G(p_{4}-p_{5},m_{f})\\
 \nonumber& &
\overline{u}_{f}(p_{3})\eslash(p_{1})(\pslash_{3}-\pslash_{1}+m_{f})
\eslash(p_{2})(\pslash_{4}+\pslash_{5}-m_{f})\gamma_{5}v_{f}(p_{4})
\end{eqnarray}
The amplitudes for the diagrams with the interchange of the two
incoming photons can be directly obtained by interchanging
$p_1,p_2$ in above amplitudes. Here, the subindex $f=t,b$, the
coefficent
\begin{eqnarray}
C_t=-\frac{16\sqrt{2}}{9}\frac{m_{t}tan\beta}{v_{w}}M_{Z}^{2}
G_{F}c_{w}^{2}s_{w}^{2}(1-\varepsilon)\\
C_b=-\frac{4\sqrt{2}}{9}\frac{m_{b}^*tan\beta}{v_{w}}M_{Z}^{2}
G_{F}c_{w}^{2}s_{w}^{2}.
\end{eqnarray}
and $G(p,m)=\frac{1}{p^{2}-m^{2}}$ is the propagator of the
particle,
$s_{w}^{2}=sin^{2}\theta_{w},c_{w}^{2}=cos^{2}\theta_{w}$
($\theta_{w}$ is the Weinberg angle)

With above amplitude, we can directly obtain the cross section
$\hat{\sigma}(\hat{s})$ for the subprocess
$\gamma\gamma\rightarrow f\overline{f}\Pi_{t}^{0}$, the total
cross section at the $e^+e^-$ linear collider can be obtained by
folding $\hat{\sigma}(\hat{s})$ with the photon distribution
function which is given in Ref\cite{Gjikia}
\begin{equation}
\sigma_{tot}(s)=\int^{x_{max}}_{x_{min}}dx_{1}\int^{x_{max}}_{x_{min}
x_{max}/x_1}dx_{2} F(x_{1})F(x_{2})\hat{\sigma}(\hat{s})
\end{equation}
where $s$ is the c.m. energy squared for $e^+e^-$ and the
subprocess occurs effectively at $\hat{s}=x_1x_2s$, and $x_i$ are
the fraction of the electrons energies carried by the photons. The
explicit form of the photon distribution function $F(x)$ is
\begin{eqnarray}                                                    
\displaystyle F(x)=\frac{1}{D(\xi)}\left[1-x+\frac{1}{1-x}
-\frac{4x}{\xi(1-x)}+\frac{4x^2}{\xi^2(1-x)^2}\right],
\end{eqnarray}
with
\begin{eqnarray}                                                    
\displaystyle
D(\xi)&=&\left(1-\frac{4}{\xi}-\frac{8}{\xi^2}\right)
\ln(1+\xi)+\frac{1}{2}+\frac{8}{\xi}-\frac{1}{2(1+\xi)^2},\\
x_{max}&=&\frac{\xi}{1+\xi}, \ \   \
\xi=\frac{4E_0\omega_0}{m^2_e}.
\end{eqnarray}
where $E_0$ and $\omega_0$ are the incident electron and laser
light energies. To avoid unwanted $e^+e^-$ pair production from
the collision between the incident and back-scattered photons, we
should not choose too large $\omega_0$. This constrains the
maximum value for $\xi=2(1+\sqrt{2})$. We obtain
\begin{eqnarray}
x_{max}=0.83, \  \   \      \  D(\xi)=1.8
\end{eqnarray}
The minimum value for x is then determined by the production
threshold,
\begin{eqnarray}
x_{min}=\frac{\hat{s}_{min}}{x_{max}s}, \ \  \  \
\hat{s}_{min}=(2m_t+M_{\Pi})^2
\end{eqnarray}
 \noindent{\bf III The results and conclusions}

In our calculations, we take $m_{t}=174$ GeV, $m_{b}=4.9$ GeV,
$v_{t}=60 Gev$, $M_{Z}=91.187$ GeV, $s^2_w=0.23$,
$G_F=1.16639\times 10^{-5}$ $(GeV)^{-2}$. There are three free
parameters involved in the production amplitudes, i.e.,
$\varepsilon,M_{\Pi}, \sqrt{s}$(for $t\bar{t}\Pi^0_t$ production)
 and $k, M_{\Pi}, \sqrt{s}$(for $b\bar{b}\Pi^0_t$ production).
To see the influence of these parameters on the production cross
section, we take the mass of toppion $M_{\Pi}$ to vary in certain
ranges 150 GeV$\leq M_{\Pi}\leq$ 350 GeV,
$\varepsilon=0.03,0.06,0.1$ and $k=0.3,0.7$ respectively.
Considering the
 center-of-mass energies $\sqrt{s}$ in planned $e^+e^-$ linear colliders
 (for example:
 TESLA), we take $\sqrt{s}$=500 GeV, 800 GeV, 1600
 GeV, respectively(but for $t\bar{t}\Pi^0_t$ production, $\sqrt{s}$=500 GeV
 is too low to produce $t\bar{t}\Pi^0_t$).

In Fig.2, taking $\sqrt{s}$=800 GeV  and
$\varepsilon=0.03,0.06,0.1$ respectively, we show the total cross
section of $t\overline{t}\Pi_{t}^{0}$ production as a function of
$M_{\Pi}$. We can see that the cross section falls sharply as the
$M_{\Pi}$ increasing and the maximum of the cross section reach
the level of 0.1 $fb$.  The phase space is depressed strongly by
large $M_{\Pi}$. For $\sqrt{s}$=1600 GeV, the results of the cross
section is shown in Fig.3. The results shown that the large
$\sqrt{s}$ can enhance the cross section significantly and the
value of the cross section is at the level of a few $fb$. This
means that large $\sqrt{s}$ is favorable for detecting $\Pi_t^0$.

The another associated production of $\Pi_t^0$ is $\gamma\gamma
\rightarrow b\overline{b}\Pi^0_t$. The results are shown in
Fig.4-5. We find that the cross section of $b\bar{b}\Pi^0_t$
production is at least two orders of magnitude smaller than that
of $t\bar{t}\Pi^0_t$ production. This is because the coupling of
$b\bar{b}\Pi^0_t$ is much smaller than the coupling of
$t\bar{t}\Pi^0_t$. So, it is difficult to detect $\Pi_t^0$ via the
process $\gamma\gamma \rightarrow b\overline{b}\Pi^0_t$. We will
not discuss the results of $\gamma\gamma \rightarrow
b\overline{b}\Pi^0_t$ in detail.

 Now, we will
focus on considering how to detect $\Pi_t^0$ via the process
$\gamma\gamma \rightarrow t\overline{t}\Pi^0_t$. It can be
concluded that the high c.m. energy is needed in order to enhance
the production rate and produce enough signals. On the other hand,
we should find the best channel to detect $\Pi^0_t$. The possible
decay modes of $\Pi^0_t$ are: $t\bar{t}$(if
$\Pi^0_t>2m_t$),$t\bar{c}, b\bar{b},gg,\gamma\gamma,Z\gamma$. For
$\Pi^0_t>2m_t$, the main decay mode is $\Pi^0_t \rightarrow
t\bar{t}$. As it is known, the couplings of toppion  to the three
families fermions are non-universal and therefore do not posses a
GIM mechanism, this non-universal feature results in a large
flavor changing coupling of neutral toppion to top and charm. So,
the decay branching ratio $Br(\Pi^0_t\rightarrow t\bar{c})$ is the
largest one when $t\bar{t}$ channel is forbidden. In SM, the cross
section of the processes with $t\bar{c}$ production should be very
small because there is no tree level flavor-changing neutral
current(FCNC) in SM. Therefore, $\gamma\gamma\rightarrow
t\bar{t}\Pi^0_t \rightarrow t\bar{t}(t\bar{c})$ is the most ideal
channel to detect neutral toppion with the clean background in SM.
Taking $M_{\Pi}=160$ GeV, we can easy get the branching ratio of
$\Pi_t^0 \rightarrow t\bar{c}$ as $66\%$. The cross section of
$e^+e^-\rightarrow \gamma\gamma\rightarrow t\bar{t}\Pi^0_t
\rightarrow t\bar{t}(t\bar{c})$ is about 4.2 fb for
$\sqrt{s}=1600$ GeV and $\varepsilon=0.06$. There are about 2000
signals can be produced via $t\bar{c}$ channel with annually
integral luminosity of $500 fb^{-1}$ at the TESLA. Such sufficient
signals can be easily detected with the clean background.

In summary, we have studied a associated production of neutral
toppion with a pair of heavy quarks in $\gamma\gamma$ collisions
in TC2 model. We find that the production rate for
$t\overline{t}\Pi_{t}^{0}$ in $\gamma\gamma$ collisions is at the
level of serval fb with high c.m. energy(for example:
$\sqrt{s}=1600$ GeV). With such production rate and clean SM
background, $\Pi_t^0$ should be dectectd experimentally at the
TESLA. But due to heavy $t\bar{t}$ pair, it is difficult to find
$\Pi_t^0$ for $\sqrt{s}=800$ GeV, except for light $\Pi_t^0$.

\newpage
\begin{figure}[ht]
\begin{center}
\begin{picture}(250,200)(0,0)
\put(-80,-70){\epsfxsize 140 mm \epsfbox{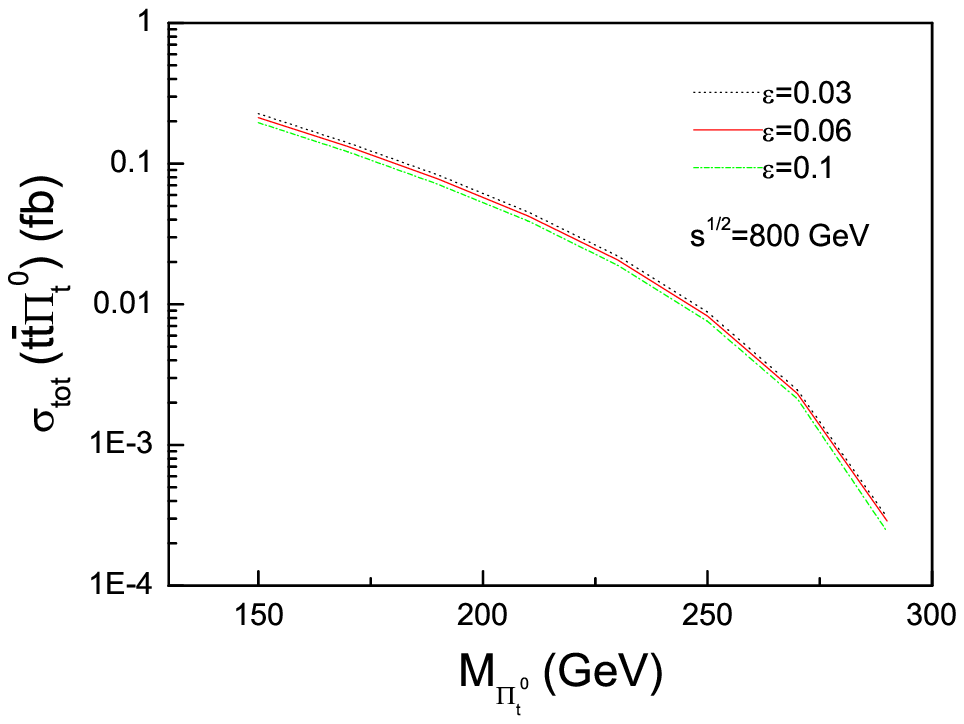}}
\put(-10,-70){Fig.2.  The total cross sections of
$t\bar{t}\Pi_t^0$ production versus the toppion }
\put(-30,-90){mass $M_{\Pi}$ for the center-mass-energy
$\sqrt{s}=800$ GeV
           and $\epsilon=0.03, 0.06, 0.1$.}
\end{picture}
\end{center}
\end{figure}

\begin{figure}[hb]
\begin{center}
\begin{picture}(250,200)(0,0)
\put(-80,-180){\epsfxsize 140 mm \epsfbox{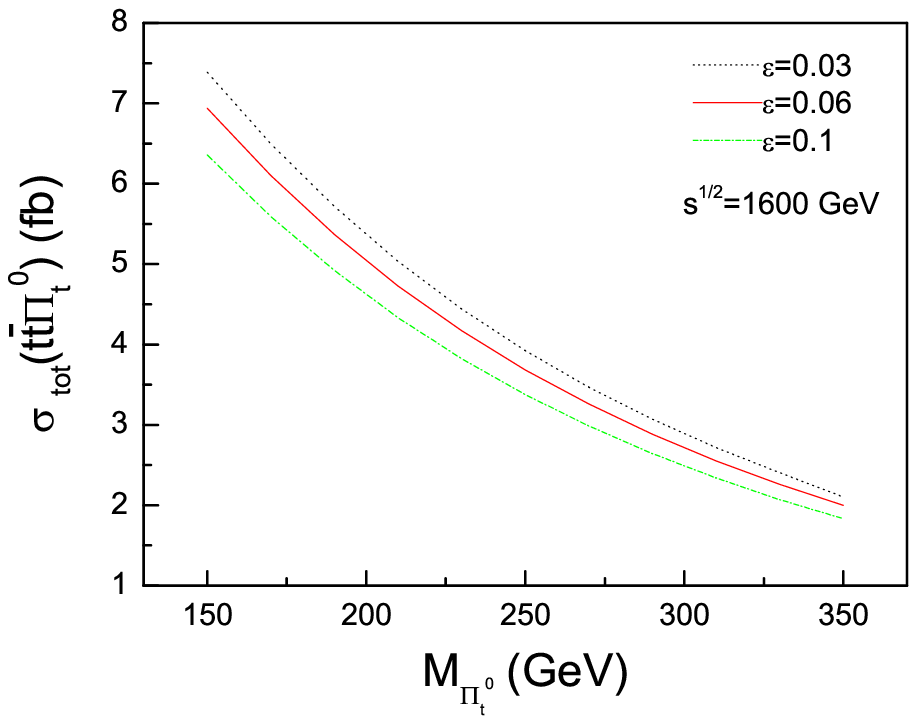}}
\put(-10,-180){Fig.3. The same plots as Fig.2 but for
$\sqrt{s}=1600$ GeV}
\end{picture}
\end{center}
\end{figure}

\newpage
\begin{figure}[ht]
\begin{center}
\begin{picture}(250,200)(0,0)
\put(-80,-70){\epsfxsize 140 mm \epsfbox{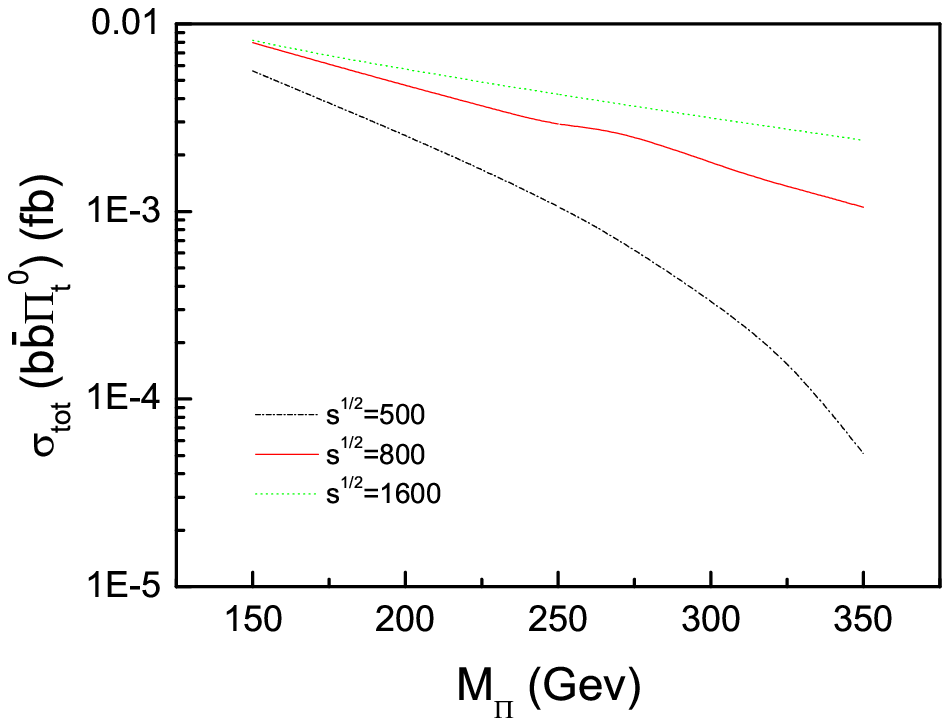}}
\put(-10,-70){Fig.4. The total cross sections of $b\bar{b}\Pi_t^0$
production versus }
\put(-30,-90){the toppion mass $M_{\Pi}$ for
$k=0.3$ and $\sqrt{s}=500,800,1600 GeV$}
\end{picture}
\end{center}
\end{figure}

\begin{figure}[hb]
\begin{center}
\begin{picture}(250,200)(0,0)
\put(-80,-180){\epsfxsize 140 mm \epsfbox{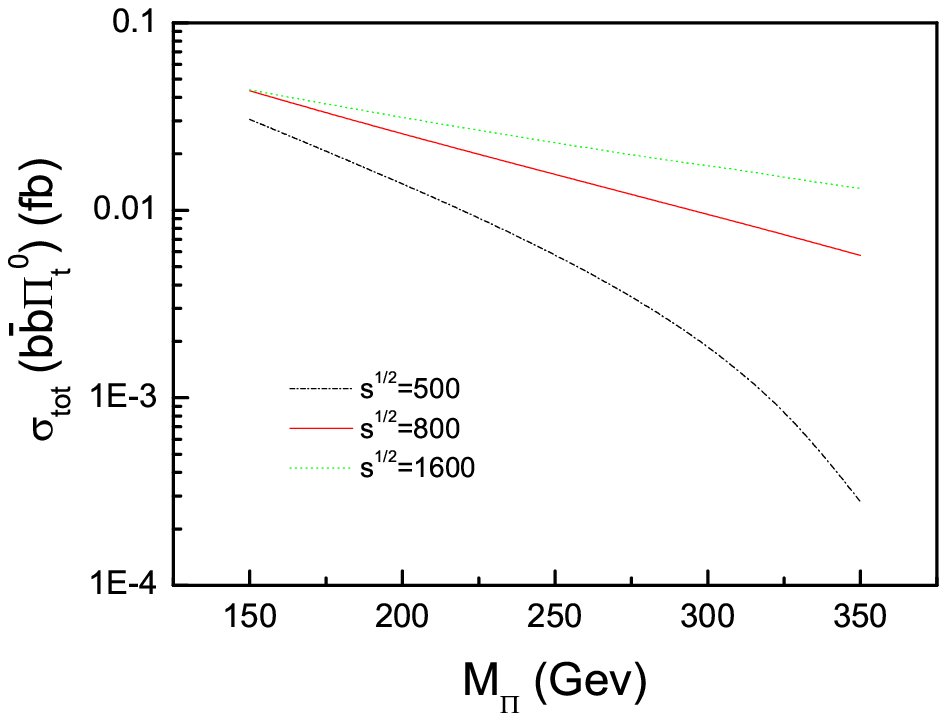}}
\put(10,-180){Fig.5. The same plots as Fig.4 but for $k=0.7$}
\end{picture}
\end{center}
\end{figure}

\end{document}